\documentclass[superscriptaddress,twocolumn,pre]{revtex4}
\usepackage{ifthen}
\newboolean{pnas}
\setboolean{pnas}{false}

\usepackage{amsmath}
\usepackage{amsfonts}
\usepackage{amssymb}
\usepackage{mathtools}
\usepackage{graphicx}
\usepackage[T1]{fontenc}
\usepackage[utf8]{inputenc}
\graphicspath{{images/}}
\usepackage{color}
\usepackage[pdfstartview=FitH,
            breaklinks=true,
            bookmarksopen=false,
            bookmarksnumbered=true,
            colorlinks=true,
            linkcolor=black,
            citecolor=black,
            urlcolor=black,
            pdftitle={},
            pdfauthor={Andreas Mayer},
            pdfsubject={}
            ]{hyperref}

\newcommand{\ud}{\mathrm{d}}

\def\(({\left(}
\def\)){\right)}                       
\def\[[{\left[}
\def\]]{\right]}

\begin{document}

\title{Regulation of T cell expansion by antigen presentation dynamics}
\author{Andreas Mayer}
\affiliation{Lewis-Sigler Institute for Integrative Genomics, Princeton University}
\author{Yaojun Zhang}
\affiliation{Princeton Center for Theoretical Science, Princeton University}
\author{Alan S. Perelson}
\affiliation{Theoretical Biology and Biophysics, Los Alamos National Laboratory}
\author{Ned S. Wingreen} 
\affiliation{Lewis-Sigler Institute for Integrative Genomics, Princeton University}
\affiliation{Department of Molecular Biology, Princeton University}
\date{\today}

\begin{abstract}
An essential feature of the adaptive immune system is the proliferation of antigen-specific lymphocytes during an immune reaction to form a large pool of effector cells. This proliferation must be regulated to ensure an effective response to infection while avoiding immunopathology. Recent experiments in mice have demonstrated that the expansion of a specific clone of T cells in response to cognate antigen obeys a striking inverse power law with respect to the initial number of T cells.
Here, we show that such a relationship arises naturally from a model in which T cell expansion is limited by decaying levels of presented antigen. The same model  also accounts for the observed dependence of T cell expansion on affinity for antigen and on the kinetics of antigen administration.
Extending the model to address expansion of multiple T cell clones competing for antigen, we find that higher affinity clones can suppress the proliferation of lower affinity clones, thereby promoting the specificity of the response. Employing the model to derive optimal vaccination protocols, we find that exponentially increasing antigen doses can achieve a nearly optimized response.
We thus conclude that the dynamics of presented antigen is a key regulator of both the size and specificity of the adaptive immune response.
\end{abstract}

\maketitle

\section{Introduction}

During an immune reaction, antigen-specific lymphocytes proliferate multiple times to form a large pool of effector cells.
T cell expansion must be carefully regulated to ensure efficient response to infections while avoiding immunopathology.
One challenge to regulation is that the number of naive T cells specific to any given antigen is biased by the VDJ recombination machinery and thymic selection \cite{Jenkins2012}, and the total number of T cells specific to an antigen further depends on previous infections with the same or similar pathogens \cite{Farber2014}. 
As a result, there are large variations in the number of precursor cells prior to an immune response.
How does the adaptive immune system properly regulate its lymphocyte expansion given varying precursor numbers? 

\begin{figure*}
    \centering
    \includegraphics{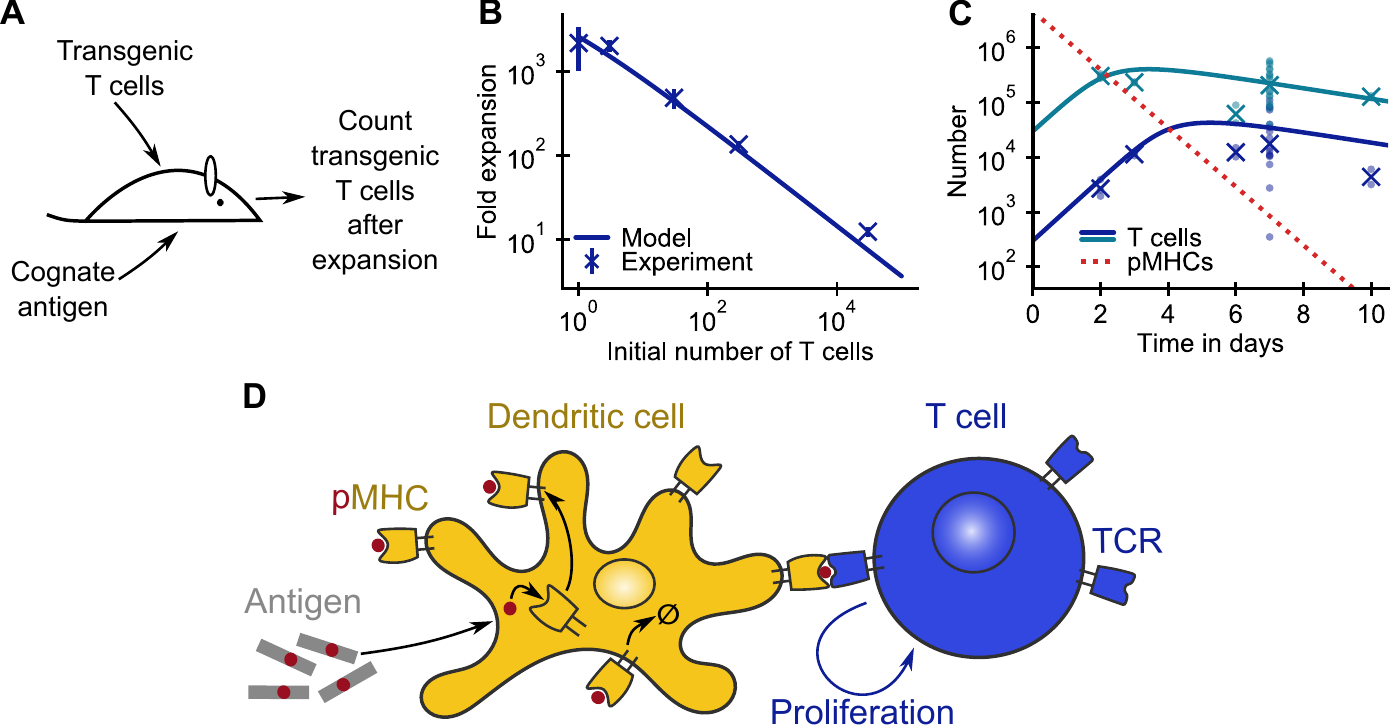}
    \caption{
        Limitation of T cell expansion by antigen decay can explain the power-law dependence of fold expansion on the initial number of cognate T cells.
        (A) Transfer of transgenic T cell clones into recipient mice allows monitoring T cell proliferation {\it in vivo} in response to cognate antigen stimulation \cite{Zehn2009,Quiel2011}. (B,C) Comparison of experimental data from \cite{Quiel2011} with model predictions.
    (B) Factor of expansion at day 7 as a function of the number of precursor T cells (crosses: geometric mean, error bars: $\pm$ SE). 
    (C) T cell and pMHC number versus time for 300 and 30,000 initial transgenic T cells (crosses: geometric mean, dots: individual mice). 
(D) Schematic of the model.
    Dendritic cells take up antigens, process them into short peptides, and present these on major histocompatibility complexes (MHCs) on their surfaces.
    T cells bind to peptide-MHC complexes (pMHC) via T cell receptors (TCRs).
    Recognition of cognate pMHC stimulates T cells to proliferate and continual antigen stimulation is needed to maintain proliferation.
    Turnover of pMHCs leads to decay of presented peptides over time.
    Fitted model parameters and asymptotic standard errors: $\alpha=1.5\pm0.3$/day, $\mu = 1.2\pm0.5$/day, $\delta=0.22\pm0.21$/day, $\log_{10} C(0) = 6.7\pm1.1$. We fixed $K=0.0$ (upper bound from fit $\simeq 700$).
    \label{figfoldexpansion}
    }
\end{figure*}

In a recent study by Quiel et al. \cite{Quiel2011}, transgenic CD4$^+$ T cells specific for a peptide from cytochrome C were adoptively transferred into mice. The number of transgenic T cells was then tracked in response to immunization with cytochrome C (Fig.~\ref{figfoldexpansion}A). It was found that fold expansion of the transgenic T cells depends as an inverse power law on the number of transferred precursor cells (Fig.~\ref{figfoldexpansion}B). Here, we propose a simple mathematical model to describe the dynamics of T cell expansion. We show that the power-law behavior arises naturally when T cell proliferation is limited by decaying antigen availability and we predict how the power-law exponent is related to the rates of lymphocyte proliferation/death and of antigen decay. Our proposal adds to previous modeling efforts to explain the surprising power-law relation in terms of negative feedback regulation \cite{Bocharov2011} or the grazing of peptide-major histocompatibility complexes (pMHCs) by T cells \cite{DeBoer2013}. We extend our analysis of the model to show that it can also explain how T cell expansion depends on antigen affinity \cite{Zehn2009} and on the the kinetics of antigen administration \cite{Johansen2008}. We thus identify the dynamics of presented antigen as a key regulator of the size of an immune response.

\section{Results}

\subsection{A simple model explains power-law dependence of fold expansion on precursor number}

We propose a minimal model of T cell expansion in which T cell proliferation is regulated by the level of presented cognate antigen (Fig.~\ref{figfoldexpansion}D). Our model has two experimentally motivated features: T cells proliferate at a saturated rate at high pMHC concentrations \cite{Quiel2011} and pMHC turnover leads to decaying antigen levels over time \cite{Zehn2004,Obst2005}.
We describe the dynamics of the number of specific T cells, $T$, and cognate pMHCs, $C$, using the equations
\begin{align}
    \frac{\ud T}{\ud t} &= \alpha \frac{T C}{K+T+C} - \delta T \label{eqdynTcomp}, \\
    \frac{\ud C}{\ud t} &= - \mu C \label{eqdynCcomp}.
\end{align}
Eq.~\ref{eqdynTcomp} follows from the assumption that T cells are stimulated to divide at a maximum rate $\alpha$ by binding to pMHCs.
However, the actual rate of T cell proliferation depends on how many cognate antigens are presented and on a saturation constant $K$ related to the affinity between the T cell receptor and the pMHC.
The form of the proliferation term in Eq.~\ref{eqdynTcomp} arises from considering competition between T cells for binding to pMHCs \cite{DeBoer1995} (SI Text~\ref{appcompbindingsingle}).
Finally, T cells die at a rate $\delta$, and pMHC complexes decay at a rate $\mu$ from their initial number.

We fit the model to experimental data from Quiel et al. \cite{Quiel2011} on the relationship between the number of precursor transgenic CD4$^+$ T cells and fold expansion (Fig.~\ref{figfoldexpansion}B) -- here defined as the number of transgenic T cells at day 7 divided by the number of transgenic T cell at day 0 -- as well as to time courses of transgenic T cell population sizes (Fig.~\ref{figfoldexpansion}C). 
The model naturally reproduces the observed power-law dependence of fold expansion, and it also recapitulates the observed biphasic time course of T cell expansion: exponential proliferation at early times followed by a cross-over into a phase of slowly decreasing T cell numbers.
The model furthermore correctly predicts that the proliferation phase ends earlier for higher precursor numbers.

Given the simplicity of the model we can readily understand the power-law behavior.
For small times the amount of pMHCs is saturating for T cell binding, $C(t) \gg K$ and $C(t) \gg T(t)$, so that T cells proliferate exponentially, $T(t) \simeq T(0) e^{(\alpha-\delta) t}$.
This exponential proliferation proceeds until the pMHC concentration $C(t) \simeq C(0) e^{-\mu t}$ decays below the saturation level.
This occurs at a transition time $t^\star$ when one of the two conditions for saturation is no longer fulfilled, i.e. when $C(t^\star) \approx T(t^\star)$, or when $C(t^\star) \approx K$. The former case implies a slowdown of expansion due to competition, the latter a slowdown due to the limited affinity of the T cell for the antigen (with a crossover between the two as shown in Fig.~\ref{figfeT0andK}).  Here, from our fitted parameters we infer the system is in the competition-limited regime (the effects of affinity are discussed in the next subsection), so that the characteristic time $t^\star$ is set by $C(t^\star) \approx T(t^\star)$, from which we have
\begin{equation} \label{eqtstar}
    t^\star = \frac{1}{\alpha - \delta + \mu} \log \frac{C(0)}{T(0)}.
\end{equation}
After this time, proliferation of T cells rapidly slows down as pMHC levels continue to decline. Neglecting the small additional T cell proliferation and decay beyond this time, the fold expansion is given by
\begin{equation} \label{eqTstar}
    \frac{T(t^\star)}{T(0)} \approx \((\frac{C(0)}{T(0)} \))^{(\alpha-\delta)/(\alpha-\delta+\mu)}.
\end{equation}
Using the parameter estimates in Fig.~\ref{figantigenkinetics} caption the maximum value of $t^\star$ is smaller than 7 days, so Eq.~\ref{eqTstar} provides a good estimate of the fold expansion (Fig.~\ref{figfoldexpansionstar}). 
Thus the model naturally yields an inverse power-law dependence of T cell amplification $T(t^\star)/T(0)$ on the initial T cell population $T(0)$.
Additionally, Eq.~\ref{eqtstar} can account neatly for the earlier timing of the peak population size at higher precursor numbers.

Importantly, Eq.~\ref{eqTstar} makes concrete predictions regarding how the fold expansion depends on system parameters.
For example, it predicts the dependence of the power-law exponent of fold expansion on the rates of T cell proliferation/decay and antigen decay. It also predicts a uniform increase in fold expansion if $C(0)$ is increased. This has been observed experimentally \cite[Fig.~5A]{Quiel2011}: either an increase in the antigen dose or in the number of antigen-presenting cells leads to a uniform increase of fold expansion across precursor numbers. 
Our model further predicts that a transfer of transgenic T cells after a time delay $t_{\rm delay}$ relative to antigen administration leads to a smaller fold expansion by a factor of $e^{- t_{\rm delay} \mu (\alpha-\delta)/(\alpha-\delta+\mu)}$, compared to Eq.~\ref{eqTstar} with $T(t_{\rm delay})$ in place of $T(0)$, as some of the presented antigens will already have decayed during the time delay (Fig.~\ref{figdelay}).

The parameters of the model can be readily inferred from the data. 
The rate of decay of T cell numbers after their peak sets $\delta$.
The initial rate of increase of T cell number is given by $\alpha - \delta$, which sets $\alpha$. 
The experimentally observed power-law exponent of $\approx -0.5$ implies $\alpha-\delta \approx \mu$, from which we can infer $\mu$.
Finally, we obtain an upper limit on the saturation constant $K$ from the observation that the power-law holds down to the smallest experimental precursor numbers (Fig.~\ref{figfoldexpansion}B);
with a higher $K$ the power law breaks down for the lowest precursor numbers (Fig.~\ref{figdynamicsK}).
More precisely, the condition $K \ll C(t^\star)$ implies
$    K \ll C(0)^{(\alpha-\delta)/(\alpha-\delta+\mu)} T(0)^{\mu/(\alpha-\delta+\mu)}$.

Our model formulation has been deliberately simple to highlight our proposed explanation for the power-law relation observed in \cite{Quiel2011}: namely competition among an exponentially increasing number of T cells for an exponentially decreasing number of pMHC complexes. The model can be extended in a number of ways without altering the basic behavior. First, one might ask how the results change for a more complicated dynamics of the pMHC number that includes the initial uptake and processing of antigen by the antigen presenting cells as was done in \cite{DeBoer2013}. A simple analysis shows that the power-law scaling continues to hold as long as there is no new processing of antigen into pMHC for times larger than the smallest $t^\star$ (SI Text~\ref{appinitialdyn}). Second, one might consider active depletion of antigens through their interactions with T cells, also known as T cell grazing \cite{DeBoer2013}. We can modify the dynamical equations to include the effect of grazing while keeping the other aspects of the model intact (SI Text~\ref{appgrazing}). Simulations and mathematical analysis of this modified model show that it exhibits similar dynamics and can also fit the experimental data. Both models have in common the same robust mechanism for generating a power law: T cells proliferate at a constant rate until the exponential decay of cognate pMHCs causes proliferation to cease.

\subsection{Dependence of T cell expansion on antigen affinity}

\begin{figure}
    \centering
    \includegraphics{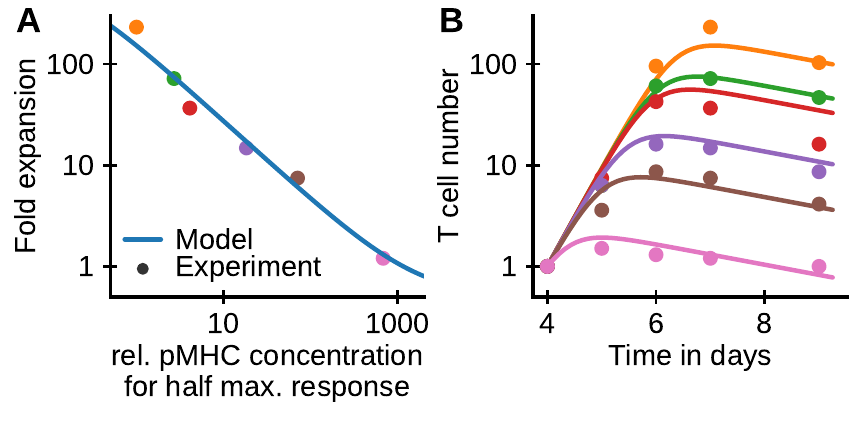}
    \caption{
        Limitation of T cell expansion by antigen decay can account for the power-law dependence of fold expansion on affinity.
        (A,B) Comparison of data from an experiment with {\it Listeria monocytogenes} strains expressing different antigens \cite{Zehn2009} with model predictions.
        (A) Factor of expansion of the transgenic T cells at day 7 relative to day 4 versus the concentration of different pMHCs needed to elicit half-maximal Interferon-$\gamma$ response from the T cells.
        (B) T cell number versus time for the different strains. 
        Fitted model parameters and asymptotic standard error: $\alpha=2.47\pm0.13$/day, $\mu = 3.1\pm0.3$/day, $\delta=0.23\pm0.06$/day, $\log_{10} C(4) = 3.22\pm0.17$, $T(4)=0.92\pm0.10$. The number of transgenic T cells are calculated from their fraction $f$ of total T cells as $f/(1-f)$ with the number at day 4 set to 1. 
    \label{figzehn}
    }
\end{figure}

T cells only respond to ligands that bind sufficiently strongly to their T cell receptor, which is the basis of the specificity of the adaptive immune system. Above this binding threshold there is a large range of possible affinities that can stimulate T cell expansion. How does the affinity of a T cell clone for the pMHC, as typically measured by a dissociation constant,  affect its expansion? In our model, affinity is captured by the parameter $K$, which is equal to the concentration of free pMHC at which the proliferation rate of the T cells is half-maximal in the non-competitive regime. Clearly, affinity is important whenever pMHC concentrations are near or below $K$. Applying our model to affinity-limited expansion in the context of a pMHC concentration that is decreasing over time makes specific predictions about how T cell proliferation depends on affinity, which we compare to experimental data from \cite{Zehn2009}. Additionally, we demonstrate that affinity can play an important role in expansion for mixtures of T cells of different affinities even if pMHC concentrations are higher than all $K$s. 

When proliferation of T cells is limited by their antigen affinity, low affinity T cells stop proliferating earlier, and consequently their fold expansion is smaller.
In contrast to the competitive regime (Eq.~\ref{eqtstar}), the characteristic time $t^\star$ at which exponential proliferation stops is now set by $C(t^\star) \approx K$. This yields a logarithmic dependence of $t^\star$ on the binding constant,
\begin{equation} \label{eqtstarK}
    t^\star = \frac{1}{\mu} \log (C(0)/K),
\end{equation}
and we find that the fold expansion scales as
\begin{equation} \label{eqTstarK}
    \frac{T(t^\star)}{T(0)} \approx \((C(0)/K\))^{(\alpha-\delta)/\mu},
\end{equation}
i.e. as an inverse power law with respect to $K$.

We compare our model predictions to data from another adoptive transfer experiment, where the effect of antigen affinity was examined \cite{Zehn2009}.
In this study the expansion of transgenic CD8$^+$ T cells was tracked in response to an infection by {\it Listeria monocytogenes}. Different strains of the bacterium were engineered to express altered ligands with different affinities for the transgenic T cells. This experimental design allowed direct study of how T cell expansion depends on antigen affinity {\it in vivo}. A reanalysis of the published data confirms our predictions that the fold expansion depends as a power law on antigen affinity (Fig.~\ref{figzehn}A, data points) and that peak population sizes are reached earlier when affinity is lower (Fig.~\ref{figzehn}B, data points).

Going beyond these qualitative predictions we can fit our model to the data of \cite{Zehn2009} quantitatively. To do so we set $K$ to the experimentally determined concentration of pMHC needed for half of T cells to show a detectable interferon-$\gamma$ response. We then fit the rates $\alpha, \mu, \delta$ as well as the day 4 pMHC concentration $C(4)$ and T cell number $T(4)$ to the time course data. In the affinity-limited regime the denominator in Eq.~\ref{eqdynTcomp} can be approximated by $K+C$, which makes the equation linear in $T$ and thus independent of the relative T cell and pMHC numbers. The fitted model closely reproduces both the observed dependence of the fold expansion on the affinity (Fig.~\ref{figzehn}A) and the full time courses of T cell numbers for different affinities (Fig.~\ref{figzehn}B).

\begin{figure}
    \centering
    \includegraphics{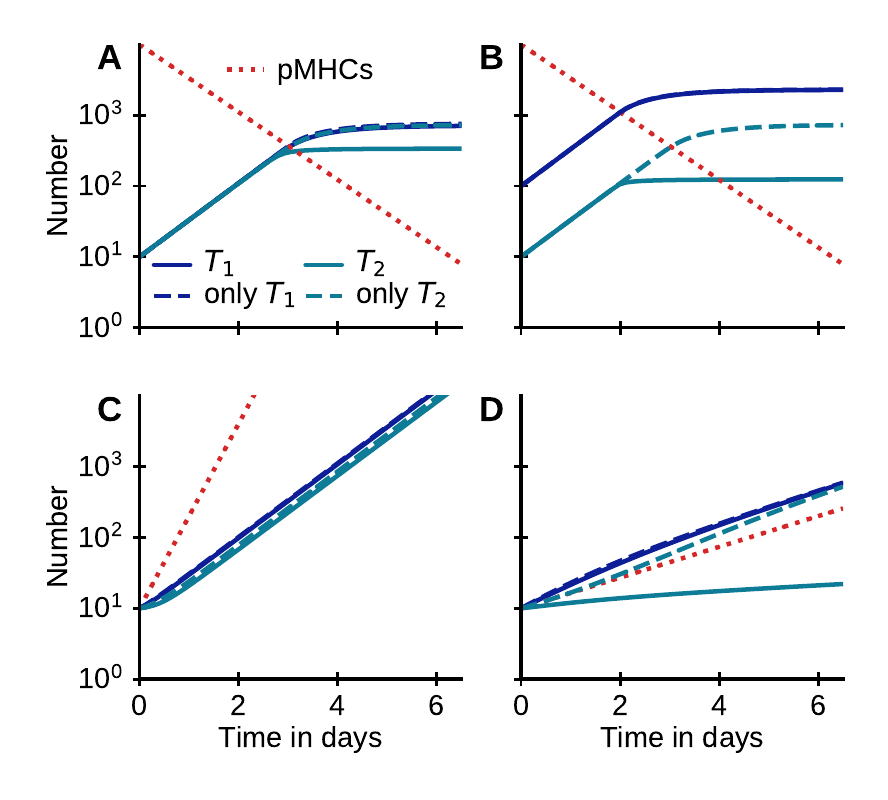}
    \caption{
        High affinity T cells can outcompete low affinity T cells for access to pMHCs, even when pMHC concentrations exceed all T cell affinities.
        (A-D) Comparisons of the time courses of expansion of mixtures of two types of T cells with high, $K_1 = 1$, and low, $K_2=10$, affinities (solid curves), to the expansion of the T cells on their own (dashed curves).
    (A,B) Proliferation driven by a large, but exponentially decreasing number of pMHCs ($\mu=1.1$) starting from equal (A) and unequal (B) initial T cell numbers.
    (C,D) Proliferation driven by a small, but exponentially increasing number of pMHCs mimicking antigen dynamics early in an infection ($C(t) \propto  e^{\gamma t}$). Competition outcome depends strongly on whether pMHC levels increase faster $\gamma = 3$ (C) or slower $\gamma = 0.5$ (D) than T cells proliferate.
Parameters: $\alpha=1.2$, $\delta = 0$.
    \label{figmultiaffinity}
    }
\end{figure}

So far, we have considered proliferation of a clone of T cells with a particular affinity for the pMHCs. However, the preinfection T cell repertoire specific to an antigen is typically broad with many different T cell clones of different affinities participating. 
How does the presence of other T cells of different affinities affect proliferation of a particular T cell clone?
Following \cite{DeBoer1995} we generalize our competitive binding model to multiple T cell populations of varying affinities (SI Text~\ref{appcompbinding}).
When pMHCs are abundant the presence of other T cells with different affinities does not change proliferation of individual clones (Eq.~\ref{eqBinocompetition}). 
However, when T cells are competing for antigens high affinity T cells enjoy preferential binding to antigen, with the magnitude of the preference set by the affinity ratio (Eq.~\ref{eqBistrongcompetition}). This implies, perhaps surprisingly, that even for pMHC concentrations well above all T cell affinities, the higher affinity T cells proliferate much faster (Fig.~\ref{figmultiaffinity}).

How differential proliferation of T cells of different affinities affects overall fold expansion depends on both the onset and extent of the competitive regime. For example, consider the dynamics of two T cell clones with relatively high and low affinities for the same pMHC antigen (Fig.~\ref{figmultiaffinity}).  If the pMHC concentration declines from a very high saturating initial level, most of the expansion happens before the onset of competition, and so the overall difference in fold expansion is low (Fig.~\ref{figmultiaffinity}A). However, high affinity T cells can still effectively shut off proliferation of a smaller population of low affinity T cells before these would stop proliferating on their own, resulting in a lowered fold expansion of the low affinity T cells (Fig.~\ref{figmultiaffinity}B).
In the experiments of Zehn et al. \cite{Zehn2009} the transgenic T cell clones expanded to comparable numbers at day 4 when stimulated by antigens with different affinities (Fig.~\ref{figzehn}B), which suggests that affinity did not play a major role in the first 4 days of expansion. In our model, initial T cell expansion is independent of affinity as long as the pMHC concentration rises quickly to levels which saturate proliferation for all T cells (Fig.~\ref{figmultiaffinity}C). By contrast, if pMHC concentration rises more slowly than the maximum rate of T cell proliferation, high affinity T cells can outcompete low affinity T cells for pMHCs and thereby suppress expansion of the low affinity T cells (Fig.~\ref{figmultiaffinity}D). 

In summary, the same model used to explain the power-law dependence of expansion on initial cell numbers for CD4$^+$ T cells is also able to quantitatively explain the effect of affinity on CD8$^+$ T cell expansion. The two experiments reveal limiting cases of the model, competition-limited versus affinity-limited. In between these limiting cases, the effects of antigen affinities, T cell numbers, and pMHC concentrations combine in interesting ways to differentially influence T cell expansion.

\subsection{Dependence of T cell expansion on antigen kinetics}

\begin{figure}
    \centering
    \includegraphics{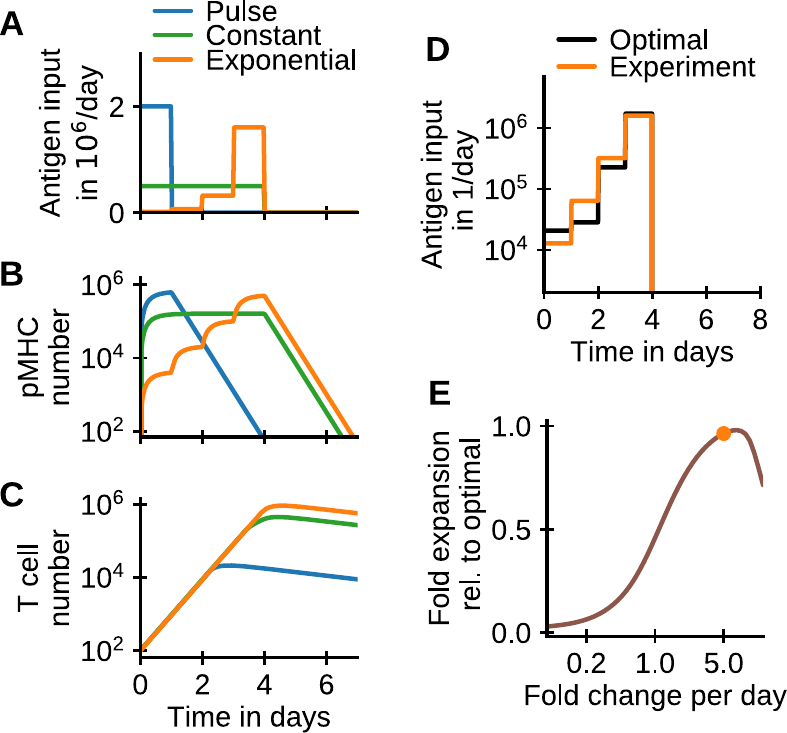}
    \caption{Impact of antigen kinetics on T cell proliferation.
    (A) Antigen input schedules: single pulse, constant input, and exponentially increasing input (increase each day by 5 fold as in \cite{Johansen2008}).
    (B) Dynamics of pMHCs.
    (C) Dynamics of T cells.
    (D) The optimal schedule is close to the experimentally used exponential schedule. The antigen input schedule over 4 days that optimizes fold expansion at day 6 was computed numerically using a projected gradient algorithm \cite{Mayer2015}.
    (E) Fold expansion for exponential schedules as a function of the fold increase per day, with the experimental schedule indicated by the dot.
    Parameters: $\alpha$, $\mu$, $\delta$ as in Fig.~\ref{figzehn}, $K = 10$, $C(0) = 0, T(0) = 100$, total administered antigen $2 \cdot 10^6$.\\
    \label{figantigenkinetics}
    }
\end{figure}

A recent experimental study with mice \cite{Johansen2008} has shown that spreading vaccine antigen administration into multiple, smaller doses instead of a single, larger dose can increase CD8$^+$ T cell responses. Is our model consistent with this observed dependence of T cell expansion on antigen dosing?

To address this question we add a time-varying antigen input rate $\nu(t)$ to Eq.~\ref{eqdynCcomp} (see Materials and Methods Eq.~\ref{eqdynC}).
We simulate the system of equations for three cases with equal total antigen input but different schedules (Fig.~\ref{figantigenkinetics}A): (1) a single antigen pulse of 1-day length, (2) constant input over 4 days, and (3) exponentially increasing input over 4 days.
These inputs lead to markedly different pMHC levels over time (Fig.~\ref{figantigenkinetics}B). To mimic the experimental protocol of \cite{Johansen2008}, we analyze the impact of these different pMHC kinetics on T cell numbers at day 6 (Fig.~\ref{figantigenkinetics}C).
We find that the exponentially increasing input leads to the largest T cell expansion (orange curve), ahead by a factor of 2 relative to constant antigen dosing (green curve), and the single shot protocol (blue curve).
While the experiment shows a larger fold-advantage of the exponential protocol over the constant protocol, our model correctly predicts the observed order of response amplitudes.
Our model provides a simple explanation for the different potencies of the protocols. Early on when T cell numbers are very low, even low pMHC concentrations are sufficient for all T cells to proliferate at their maximal rate. Later on, as T cell numbers increase, competition of T cells for pMHC can limit their expansion. Ideally, pMHC levels should rise in tandem with T cells to minimize competition and maximize stimulation at all times. By contrast, dosing schedules with high levels of pMHCs at early times are wasteful of the finite antigen budget: While initial pMHC levels are higher than needed for full T cell stimulation, these levels decay to become insufficient to stimulate the rising T cell population at later times. The exponentially increasing input protocol better synchronizes antigen levels with T cell proliferation and therefore leads to a higher fold expansion. Indeed, we find that an optimized daily-dose protocol is close to the experimentally used exponential protocol (Fig.~\ref{figantigenkinetics}D) using the derived parameters for CD8+ T cells from Fig.~\ref{figzehn}. It is worth noting that experimental choice of 5-fold increases of dose per day \cite{Johansen2008} is surprisingly close to optimal among exponential protocols (Fig.~\ref{figantigenkinetics}E).

\section{Discussion}

Prompted by the surprising observation of a power-law dependence of T cell fold expansion on initial cell numbers \cite{Quiel2011}, we developed a simple mathematical model in which T cell proliferation is stimulated by a dynamically changing number of cognate pMHC molecules, and showed that it quantitatively yields the observed power law. We then explored more generally how T cell numbers, TCR affinity for antigen, and the dynamics of pMHC presentation combine to regulate T cell expansion in different regimes. Testing these results against other experimental data, we found that the model correctly predicts a power-law dependence of T cell expansion on affinity for lower affinity antigens \cite{Zehn2009}, and also predicts the observed enhanced efficacy of stretching a fixed total antigen dose over several days \cite{Johansen2008}.
We further explored competition among multiple T cell clones with differing affinities, and generated testable predictions for future experiments.

The core of our model is that T cell expansion is regulated by dynamically changing levels of presented antigens.
The dynamics of pMHCs is assumed to be characterized by a fast processing of antigens by antigen-presenting cells followed by a slower decay.
The first assumption of rapid processing seems well justified for the subcutaneous injection of antigens employed in \cite{Quiel2011}, but might be more questionable in \cite{Zehn2009}, where live replicating bacteria are used, as this might lead to continued processing of new antigens by dendritic cells. However, as the bacterial load declines rapidly after reaching a peak at three days post-infection \cite{Pamer2004,Wang2011}, newly generated pMHCs likely play a small role compared to the turnover of the already presented pMHCs at the late stages of infection analyzed in Fig.~\ref{figzehn}.
The second assumption of the decay of presented pMHCs has a known mechanistic basis in the turnover of MHCs by ubiquitylation \cite{Roche2015} and in the apoptosis of activated antigen presenting cells \cite{Kushwah2010}.
Furthermore, our inferred decay constants
are within the range of decay constants reported for different antigens bound to dendritic cells
\cite{Zehn2004} and are also roughly compatible with direct measurements of the decrease of the stimulatory capacity of antigen-presenting cells in transgenic mice after switching off inducible antigen production \cite{Obst2005}. 
One limitation of our model, particularly for replicating antigens, is that we have neglected any influence of the epitope specific pMHC density on the surface of an antigen presenting cell, which may also play a role in determining T cell stimulation and expansion \cite{Wherry1999}.

The kinetics of antigen administration has been shown to influence the magnitude of T cell \cite{Johansen2008} and B cell \cite{Tam2016} immune responses. This finding has implications for the rational design of vaccination strategies \cite{Bachmann2010}, but open questions remain regarding how to optimize dosing for high magnitude of response and/or high affinity of the responding cells \cite{Tam2016}. Our modeling suggests that exponentially increasing doses are close to optimal for maximizing the magnitude of T cell response. We further find that selection for higher affinity is strongest when T cells compete for antigen stimulation.
Selection for affinity is thus predicted to be more stringent for lower or more slowly increasing antigen levels and for larger or faster growing prior T cell populations. This could have important implications for patterns of immunodominance in primary versus secondary infections: In secondary infections competition is expected to be stronger as preexisting memory cells specific to the pathogen are usually present in higher numbers and can also proliferate faster.

Looking ahead, our model could be further extended to make it more realistic.
Including single-cell stochasticity could help clarify how reproducible population-level expansion arises despite stochasticity at the single-cell level \cite{Hawkins2007} and would also allow connections to recent experimental studies, that have revealed substantial heterogeneity of the immune responses of single-cells \cite{Buchholz2018}. 
Furthermore, the model could be extended to account for the diverse compartments (different lymph nodes, the spleen, different tissues) \cite{Farber2014} and T cell subtypes \cite{Bocharov2011} involved in an immune response. Spatial or cellular heterogeneity can create separate niches in which T cells compete for proliferation and survival, which may provide a further layer of regulation of T cell expansion. 

Among the open questions raised in our study we highlight two:
First, other than regulating T cell expansion during an acute infection, how else might antigen presentation levels influence T cell populations? 
Do they play a role in thymic selection or the dynamics of naive T cells competing for self-antigens?
Specifically, recent work in ecology \cite{Posfai2017} suggests that T cell grazing, which consumes antigens, could allow for the coexistence of a diverse naive repertoire despite competition for self-antigens.
Second, regulation of T cell population dynamics can be achieved by tuning of system parameters. For example, smart control of the lifetime of presented pMHC complexes could induce the “right” amount of T cell amplification. Might system parameters have evolved to be close to optimal and/or could some parameters also be regulated during an immune response to provide a robust response to infections while avoiding autoimmunity?

\section{Materials and Methods}

\paragraph{Experimental studies.} We used previously published data from \cite{Quiel2011} and \cite{Zehn2009}. In both studies transgenic T cells were transferred into recipient mice and their expansion in response to antigenic stimulation was subsequently studied.
Quiel et al. \cite{Quiel2011} studied the response of 5C.C7 CD4$^+$ T-cells to a subcutaneous administration of pigeon cytochrome C and lipopolysaccharide. The number of transgenic T cells was determined by quantitative real-time PCR calibrated against a reference with known T cell numbers.
Zehn et al. \cite{Zehn2009} used OT-1 CD8$^+$ T cells specific to an ovalbumin peptide. They injected into the mice {\it Listeria monocytogenes} expressing either wild type ovalbumin or altered ovalbumin with specific amino acid substitutions in the antigenic region. The percentage of transgenic T cells within the spleen of mice was quantified by flow cytometry to measure the relative expansion of transferred versus endogeneous cells.
To assess T cell affinity for the different ligands they were stimulated with RMA-S cells loaded with different doses of antigen for five hours. T cell responses were assessed by Interferon-$\gamma$ staining and quantified by the peptide dose needed for half-maximal response probability.

\paragraph{Model equations.} We consider deterministic birth-death type models for the dynamics of the T cell population, $T$, and the concentration of pMHCs, $C$, of the following form
\begin{align}
    \frac{\ud T}{\ud t} &= \alpha(T, C, K) T - \delta T \label{eqdynT}, \\
    \frac{\ud C}{\ud t} &= \nu(t) - \mu(T, C, K) C \label{eqdynC}.
\end{align}
By binding to pMHCs, T cells are stimulated to proliferate at a rate $\alpha(T, C, K)$, which can depend on the availability of antigen, competition with other T cells, and on a parameter $K$ related to the affinity of the T cell receptor for pMHC (SI Text~\ref{appcompbinding}).
T cells die at a rate $\delta$.
The uptake of antigen and its processing leads to a time-varying influx $\nu(t)$ of pMHCs. 
The pMHC complexes decay at a rate $\mu(T, C, K)$, which may depend on T cell numbers (see SI Text~\ref{appgrazing}).
In the main text we make a number of simplifying assumptions regarding $\alpha$ and $\mu$. Several more general cases are considered in the SI Text.

\paragraph{Numerical simulations.} The differential equations describing the T cell / pMHC dynamics were integrated using a Runge-Kutta method of order 4(5) due to Dormand and Prince (routine {\it dopri5} in scipy \cite{scipy}). Parameters were determined using least-square fits to log-transformed mean cell numbers weighted by the number of repeat experiments. Asymptotic standard errors of the parameters were calculated based on the residual covariance matrix. An upper bound for $K$ was defined by the value of $K$ which leads to a unit increase of the sum of squared residuals \cite{Press2007}.

{\bf Acknowledgements.}
The authors thank Grégoire Altan-Bonnet for helpful discussions, and Gennady Bocharov and Zvi Grossman for providing experimental data. This work was started from discussions at the Kavli Institute of Theoretical Physics, University of California, Santa Barbara, and supported in part by NSF Grant No. PHY-1748958, NIH Grant No. R25GM067110, and the Gordon and Betty Moore Foundation Grant No. 2919.01. This work was supported by a Lewis–Sigler fellowship (AM), the Princeton Center for Theoretical Science and the National Science Foundation grant PHY-1607612 (YZ), National Institutes of Health grants R01-OD011095 and R01-AI028433 (ASP), and the Center for the Physics of Biological Function funded by the National Science Foundation grant PHY-1734030 (NSW). Portions of this work were performed under the auspices of the U.S. Department of Energy under contract DE-AC52-06NA25396.

\section*{Significance}
Antigen-specific T cells proliferate multiple times during an immune reaction to fight against disease. This expansion of T cells must be carefully regulated to ensure an effective defense while avoiding autoimmunity. One challenge to regulation is that the initial size of antigen-specific T cell clones can be quite variable. Intriguingly, a recent study in mice found that the fold expansion of a T cell clone depends as an inverse power law on its initial size. We propose a simple mathematical model which naturally yields the observed power-law relation. Our model accounts for multiple experiments on T cell proliferation, suggests optimal vaccination protocols, and highlights the dynamics of presented antigen as a key regulator of the size of an immune response.

\clearpage

\appendix
\renewcommand\appendixname{SI Text}
\setcounter{figure}{0}
\makeatletter 
\renewcommand{\thefigure}{S\@arabic\c@figure}
\makeatother

\section{Competitive binding model}
\label{appcompbinding}

The preinfection T cell repertoire specific to an antigen is typically broad consisting of many T cell clones of varying affinities. Competition between all of these T cells for binding to a limited number of pMHC complexes can reduce T cell proliferation. To make our work self-contained we derive a mathematical framework to analyze how such competition influences T cell expansion, largely following earlier work by Perelson and DeBoer \cite{DeBoer1995} and Borghans et al. \cite{Borghans1996}.

\subsection{General framework}

We consider the following dynamical equations to describe the expansion of multiple T cell clones competing for the cognate antigen:
\begin{align}
    \frac{\ud T_{i}}{\ud t} &= \alpha B_i - \delta T_i, \label{Ti}\\
    \frac{\ud C}{\ud t} &= -\mu C. \label{C}
\end{align}
The first equation describes the dynamics of the number of T cells, $T_i$, in the $i\textsuperscript{th}$ clone, with $i$ running from $1$ to the number of clones $n$ that have significant affinity for the antigen. All T cells are assumed to proliferate at a rate $\alpha$ upon binding to pMHC complexes; thus the overall rate of proliferation of T cells in the $i\textsuperscript{th}$ clone is proportional to the number of bound T cell-pMHC complexes $B_i$. All T cells are assumed to die at a rate $\delta$. The second equation describes the dynamics of the number of cognate pMHCs, $C$. The pMHCs are presented on the surface of antigen-presenting cells and are assumed to decay at a rate $\mu$. 

The specific form of the proliferation term in Eq.~\ref{Ti} can be derived by considering the following scheme \cite{DeBoer1995} which resembles an enzymatic reaction:      
\begin{equation} \label{}
    T_{ui}+C_u  \underset{k_{di}}{\stackrel{k_{ai}}{\rightleftharpoons}} B_i \stackrel{k_{pi}}{\rightarrow} 2T_{ui}+C_u,
\end{equation}
where $T_{ui}$ is the number of unbound T cells in the $i\textsuperscript{th}$ clone, $C_u$ is the number of unbound pMHCs, and $k_{ai}$, $k_{di}$ and $k_{pi}$ are, respectively, rate constants for association and dissociation of the bound complex $B_i$ and for the proliferation of T cells in $i\textsuperscript{th}$ clone upon binding. 
After the T cell is stimulated to divide it dissociates from the APC leaving the pMHC unbound.

The dynamical equations corresponding to the above scheme are:
\begin{align}
    \frac{\ud T_{ui}}{\ud t} &= (2k_{pi}+k_{di})B_i - k_{ai}T_{ui}C_u, \\
    \frac{\ud B_i}{\ud t} &= k_{ai}T_{ui}C_u - (k_{pi}+k_{di})B_i.
\end{align}
The total number of T cells in the $i\textsuperscript{th}$ clone is the sum of the number of unbound T cells and bound complexes $T_i=T_{ui}+B_i$, so that adding the above two equations together we have
\begin{align}
    \frac{\ud T_{i}}{\ud t} = k_{pi}B_i,
\end{align}
which gives rise to the proliferation term in Eq.~\ref{Ti} after setting $k_{pi}$ to $\alpha$.  

Assuming that the binding/unbinding kinetics is much faster than the proliferation of T cells, we adopt the quasi-steady-state approximation $dB_i/dt=0$ for all the bound complexes. This leads to the following equation
\begin{equation} \label{}
T_{ui}C_u-K_iB_i=0,
\end{equation}
where $K_i=(k_{pi}+k_{di})/k_{ai}$ is the saturation constant related to the affinity between the T cell receptors in the $i\textsuperscript{th}$ clone and the pMHC. Expressing the number of unbound T cells and pMHCs in terms of the total numbers of T cells and pMHCs and of bound complexes, $T_{ui}=T_i-B_i$ and $C_u=C-\sum B_i$, we finally obtain an equation relating the number of bound complexes to the total numbers of T cells ${T_i}$ and of pMHCs $C$ and the saturation constants $K_i$:  
\begin{equation} \label{exactsol}
(T_i-B_i)(C-\sum_{j=1}^n B_j)-K_iB_i=0.
\end{equation}
The above set of equations (with $i$ running from $1$ to $n$) can be readily solved numerically to find $\{B_i\}$ given $\{T_i\}$, $C$ and $\{K_i\}$. Analytical solutions are also possible in special cases, e.g. when competition only involves one or two T cell clones or for certain limiting cases.   

\subsection{Competitive binding for a single T cell clone}
\label{appcompbindingsingle}

Following the general framework, we first consider the simplest case of competitive binding within a single population of T cells, all with equal affinity for the cognate antigen. This case can be relevant in practice. For example, in experiments a single clone of transgenic T cells and cognate antigens with high specificity for each other may be transferred together \cite{Quiel2011}, or in a natural setting one of the T cell clones in a preinfection repertoire may have a much higher specificity for a particular pathogen-derived antigen than the other clones, so that competition occurs mainly within this highly specific population of T cells.

From Eqs.~\ref{Ti} and \ref{C}, the dynamical equations for the expansion of a single T cell clone are
\begin{align}
    \frac{\ud T}{\ud t} &= \alpha B(T,C,K) - \delta T, \label{Tsingle}\\
    \frac{\ud C}{\ud t} &= -\mu C, \label{Csingle}
\end{align}
where the analytical expression $B(T,C,K)$ can be derived from the single-clone version of Eq. ~\ref{exactsol}:
\begin{equation}
    B^2 - (K + T + C) B + T C = 0.
\end{equation}
This quadratic equation has the solution
\begin{equation} \label{eqB}
    B = \frac{1}{2} \[[K + T + C - \sqrt{(K+T+C)^2 - 4 T C}\]],
\end{equation}
where the choice of the minus sign is determined by the conditions that the number of bound complexes is smaller than the total numbers of T cells and pMHCs, $B<T$ and $B<C$. When $TC \ll (K+T+C)^2$, we can expand the exact solution Eq.~\ref{eqB} with respect to the small parameter $TC/(K+T+C)^2$. This leads to a convenient approximate solution for $B$:
\begin{equation} \label{eqBapp}
    B = \frac{T C}{K+T+C}.
\end{equation}
Eq.~\ref{eqBapp} has been obtained previously in \cite{DeBoer1995}. (The largest deviation between the exact and approximate  expressions occurs when $K \to 0$ and $T=C$, where there is a factor of $2$ difference between the exact solution $B=T$ and the approximate solution $B=T/2$.) Substituting the approximate solution Eq.~\ref{eqBapp} into Eq.~\ref{Tsingle}, we obtain Eqs.~\ref{eqdynTcomp} and \ref{eqdynCcomp} in the main text for the dynamics of the T cell and pMHC numbers.

\subsection{Competitive binding for two T cell clones}
\label{twocomponent}

T cell dynamics becomes more nuanced when two T cell clones with different affinities compete for binding to the same antigen. When the antigens are abundant, all T cells can proliferate freely. However, as the level of antigen decays, competition can differentially reduce the proliferation of T cells in the different clones. The expansions of two competing T cell clones are described by Eqs.~\ref{Ti} and \ref{C} with the total number of clones $n=2$. The specific forms of the numbers of the bound complexes $B_1$ and $B_2$ in the dynamical equations can be derived from Eq.~\ref{exactsol}:
\begin{align}
(T_1-B_1)(C-B_1-B_2)-K_1B_1 &=0, \label{B1}\\
(T_2-B_2)(C-B_1-B_2)-K_2B_2 &=0. \label{B2}
\end{align}
Since the above two equations are symmetric with respect to interchange of the subscript indices $1$ and $2$, the analytical expressions for $B_1$ and $B_2$ must preserve the symmetry, i.e., $B_1=f(T_1,T_2,C,K_1,K_2)$ and $B_2=f(T_2,T_1,C,K_2,K_1)$. Therefore, we first obtain the solution for $B_1$, and the solution for $B_2$ can then be found by interchanging the subscripts $1$ and $2$ in $B_1$. Combining Eqs.~\ref{B1} and \ref{B2}, we obtain a cubic equation for $B_1$:  
\begin{align}
aB_1^3+bB_1^2+cB_1+d &=0,
\end{align}
where 
\begin{align}
a &= K_1-K_2, \nonumber\\
b &= K_1T_2+K_2T_1-(K_1-K_2)(K_1+C+T_1), \nonumber\\
c &= -T_1\[[K_1T_2+K_2T_1+K_1K_2+C(2K_2-K_1)\]], \nonumber\\
d &= K_2CT_1^2.
\end{align} 
We find the solution for $B_1$ following standard methods for solving cubic equations,  
\begin{align} 
B_1 = -\frac{1}{3a}\[[b+\text{Re}\(((-1+\sqrt{3}i)\Omega\))\]], \label{exactB1}
\end{align}
where $\text{Re}$ means real part, and
\begin{equation} 
\Omega=\sqrt[3]{\frac{Q\pm i\sqrt{4P^3-Q^2}}{2}}
\end{equation}
with $P=b^2-3ac$ and $Q=2b^3-9abc+27a^2d$. We use the plus sign in $\Omega$ when $K_2>K_1$ and the minus sign when $K_1<K_2$. 

These analytical expressions were employed to generate the results shown in Fig.~\ref{figmultiaffinity} in the main text.

\subsection{Competitive binding in limiting cases}
\label{Bisolution} 

Beyond the simplest cases of competitive binding for one or two T cell clones, analytical solutions for Eq.~\ref{exactsol} are also possible in certain limiting cases. 

In the non-competitive limit, when the number of pMHCs bound to T cells is much smaller than the total number of pMHCs, $\sum B_i \ll C$, we can neglect $\sum B_j$ in the term $C-\sum B_j$ in Eq.~\ref{exactsol} and obtain
\begin{equation} \label{eqBinocompetition}
    B_i \approx \frac{T_i C}{K_i+C}.
\end{equation}
Expressing the condition $\sum B_i \ll C$ in terms of the known parameters $C$, $T_i$ and $K_i$, the assumption underlying Eq.~\ref{eqBinocompetition} can be rewritten as $\sum T_i/(K_i+C) \ll 1$. The expression \ref{eqBinocompetition} corresponds to independent proliferation of T cells in all clones, as expected in the non-competitive limit. A less crude approximation obtained by only dropping the $B_i B_j$ terms in Eq.~\ref{exactsol} is provided in \cite{DeBoer1995}.

In the highly competitive limit, when the number of bound T cells is much smaller than the total number of T cells in each clone, $B_i \ll T_i$, we can neglect $B_i$ in the term $T_i-B_i$ in Eq.~\ref{exactsol} and obtain
\begin{equation} \label{eqBistrongcompetition}
    B_i \approx C\((1+\sum_{j=1}^n\frac{T_j}{K_j}\))^{-1}\frac{T_i}{K_i}.
\end{equation}
Expressing the condition $B_i \ll T_i$ in terms of the known parameters, the assumption underlying Eq.~\ref{eqBistrongcompetition} can be rewritten as $C/K_i \ll 1+\sum(T_j/K_j) $. Note that $C (1+\sum (T_j/K_j))^{-1}$ in Eq.~\ref{eqBistrongcompetition} is a constant prefactor for all the bound species. Therefore, the fraction of bound T cells in each clone is inversely proportional to their saturation constant. Interestingly, as long as T cell numbers
are sufficiently large that $B_i \ll T_i$, Eq.~\ref{eqBistrongcompetition} holds even when $C \gg K_i$, where naively one might have expected the competition to be affinity independent. 

\section{Influence of initial antigen dynamics}
\label{appinitialdyn}

The dynamics of antigen presentation can be more complex than simple exponential decay from an initial level. Complications include the uptake and processing of antigens by the antigen presenting cells or continued generation of new antigens, e.g. due to pathogen replication during a natural infection. How might these complications in antigen dynamics influence the power-law dependence of fold expansion on the initial T cell number? 

Here, we show that the power-law scaling continues to hold for a larger class of antigen dynamics. Specifically, we define a time $t'$ after which there is no new processing of antigens into pMHCs. The initial antigen dynamics before $t'$ can be arbitrary as long as the number of antigens remains above the saturating level for T cell proliferation. This ensures independent proliferation of T cells during the initial stages, and hence a constant ratio between $T(t')$ and $T(0)$ for all precursor numbers. Antigen dynamics beyond $t'$ will obey exponential decay $C(t)=C(t')e^{-\mu (t-t')}$ as there is no new generation of pMHCs.  Since the number of antigens remains saturating until $t'$, the transition from T cell proliferation to decay happens at a time $t^\star$ later than $t'$ for all precursor numbers. Therefore, in the competition-limited regime the power-law relation in Eq.~\ref{eqTstar} in the main text holds with $C(t')$ and $T(t')$ in place of $C(0)$ and $T(0)$,
\begin{equation}
\frac{T(t^\star)}{T(t')} \approx \((\frac{C(t')}{T(t')} \))^{(\alpha-\delta)/(\alpha-\delta+\mu)}.
\end{equation}
As $T(t')$ and $T(0)$ only differ by a constant factor, the full fold expansion preserves the inverse power-law dependence on the initial T cell number: $T(t^\star)/T(0) \propto T(0)^{-(\alpha-\delta)/(\alpha-\delta+\mu)}$.

A similar analysis also applies to the affinity-limited regime.

\section{Grazing model}
\label{appgrazing}

In the competitive binding model, the power-law scaling between the fold expansion and the initial T cell number is ensured by the termination of expansion at a fixed relative level of T cells and pMHC complexes. A similar outcome can also be achieved in a model in which T cells actively degrade pMHCs upon binding, based on the observation that T cells that bind to pMHCs have some probabilities of acquiring the pMHCs \cite{Kedl2002,Tkach2009}. This process of active acquisition has been termed {\it T cell grazing} \cite{DeBoer2013}. 

In the grazing model, the dynamics of T cells and pMHCs are described by the following equations:
\begin{align}
\frac{\ud T}{\ud t} &= \alpha \frac{T C}{K+T+C} - \delta T \label{eqdynTgrazing}, \\
\frac{\ud C}{\ud t} &= - \mu C - \beta \frac{T C}{K+T+C} \label{eqdynCgrazing}.
\end{align}
The last term in Eq.~\ref{eqdynCgrazing} describes the loss of pMHCs through grazing, where the prefactor $\beta$ is the rate of grazing upon binding. For simplicity, we assume in the following that grazing limits proliferation before competitive binding sets in, i.e. we assume $T \ll K+C$ so that the number of bound complexes can be approximated by $TC/(K+C)$. The model differs from the previous grazing model in \cite{DeBoer2013} in two major ways: In Eqs.~\ref{eqdynTgrazing} and \ref{eqdynCgrazing}, the T cell proliferation rate saturates at high pMHC concentrations, and the rate of grazing has a simple linear dependence on the T cell number. The grazing term in Eq.~\ref{eqdynCgrazing} leads to an increased loss of pMHCs at high T cell numbers. This T cell-induced loss of pMHCs is also compatible with mechanisms other than grazing. In particular, the ``grazing'' term might also be interpreted as a reduction of pMHCs through T cell-induced pMHC endocytosis followed by lysosomal degradation, which has recently been reported \cite{Furuta2012}. 

The grazing model also yields an inverse power-law dependence of the fold expansion on the initial T cell number. Here, we provide a simple explanation following our analysis of the competitive binding model. At small times, the pMHC concentration is saturating for T cell binding, $C(t) \gg K$, so that T cells proliferate exponentially $T(t) = T(0) e^{(\alpha-\delta) t}$. The initial loss of pMHCs is mainly due to natural decay, as the rate of grazing of pMHCs is proportional to T cell number, which is initially small. Mathematically this means that $C(t) = C(0) e^{-\mu t}$ as $\beta T(t) \ll \mu C(t)$. The condition $\beta T(t) \ll \mu C(t)$ holds until a characteristic time
\begin{equation}
t^\star = \frac{1}{\alpha - \delta + \mu} \log \frac{\mu C(0)}{\beta T(0)}.
\end{equation}
After this time grazing leads to a rapid decline in the pMHC level. Therefore T cell proliferation ceases and the total number of T cells after expansion can be approximated by $T(t^\star) = T(0) e^{(\alpha-\delta) t^\star}$. The fold expansion is thus approximately 
\begin{equation}
\frac{T(t^\star)}{T(0)} \approx \(( \frac{\mu C(0)}{\beta T(0)} \))^{(\alpha-\delta)/(\alpha-\delta+\mu)},
\end{equation}
displaying an inverse power-law dependence on the initial T cell number. Given the similarity of the grazing model to the competitive binding model, it is not surprising that the grazing model can also fit the data presented in Fig.~\ref{figfoldexpansion}B,C (not shown). In both models, the exponential increase of T cell number, the exponential decay of pMHCs, and the termination of expansion at a fixed relative level of T cells and pMHC complexes are the essential ingredients leading to power laws in the fold expansion of T cells.

\clearpage
\onecolumngrid

\section*{Supplementary figures}

\begin{figure}[h]
    \centering
    \includegraphics{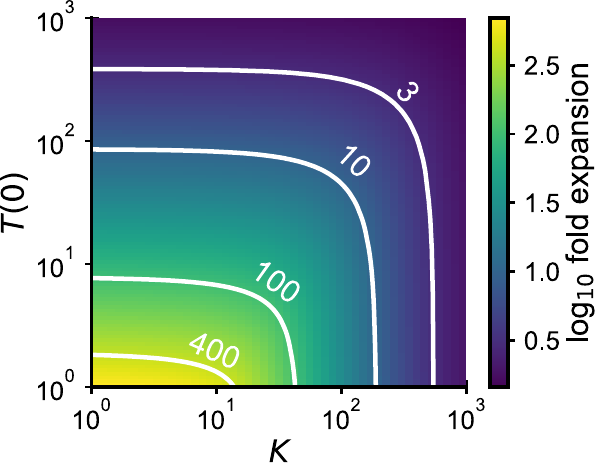}
    \caption{Dependence of fold expansion at day 7 on precursor number $T(0)$ and saturation parameter $K$. Parameters as in Fig.~\ref{figmultiaffinity}.
    \label{figfeT0andK}
    }
\end{figure}

\begin{figure}[h]
    \centering
    \includegraphics{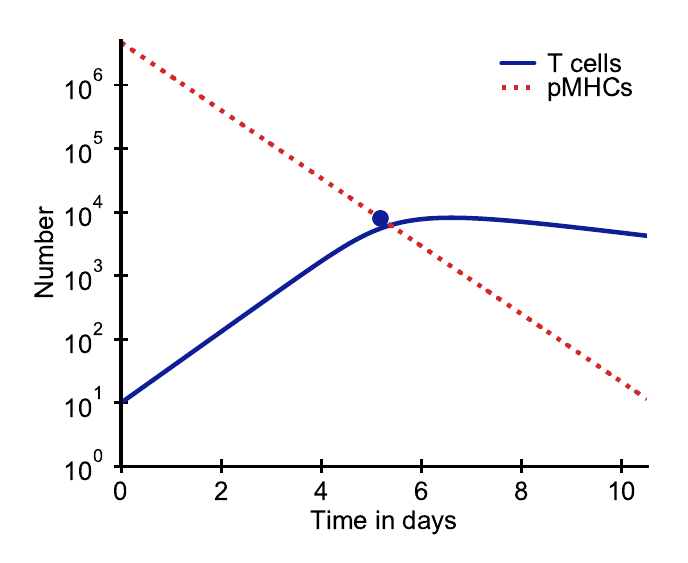}
    \caption{Modeled time courses of T cell and pMHC populations. The estimated peak time $t^\star$ and peak value $T(t^\star)$ of the T cell population as estimated by Eqs.~\ref{eqtstar} and \ref{eqTstar}, respectively, are indicated by the blue dot. Parameters as in Fig.~\ref{figfoldexpansion}. 
    \label{figfoldexpansionstar}
    }
\end{figure}

\begin{figure}[h]
    \centering
    \includegraphics{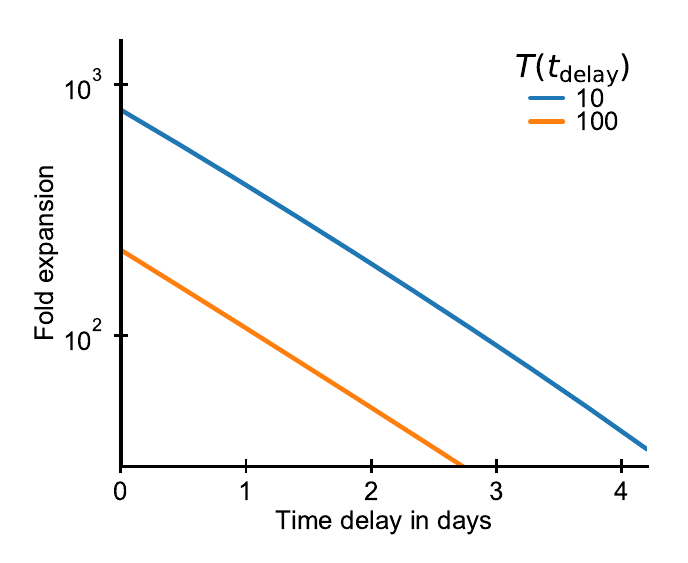}
    \caption{Transfer of transgenic T cells after a time delay relative to antigen administration is predicted to decrease fold expansion. Modeled factor of expansion at day 7 after antigen administration as a function of the time delay for initial T cell numbers 10 (blue) and 100 (orange). Parameters as in Fig.~\ref{figfoldexpansion}. 
    \label{figdelay}
    }
\end{figure}

\begin{figure}[h]
    \centering
    \includegraphics{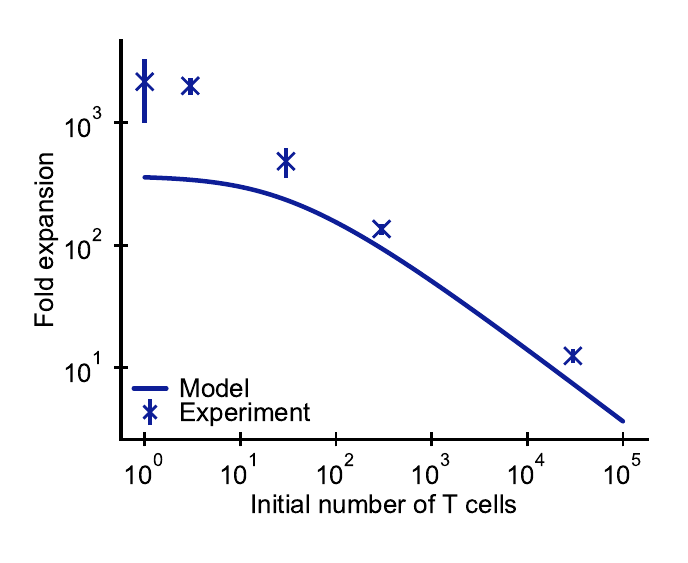}
    \caption{Modeled T cell dynamics as in Fig.~\ref{figfoldexpansion} with a saturation parameter $K = 10^{4}$ above the upper bound inferred from the data. At low precursor numbers, T cell expansion becomes affinity-limited, which caps fold expansion and leads to deviations from power-law behavior.
    \label{figdynamicsK}
    }
\end{figure}

\end{document}